\documentclass[12pt]{article}
\newcommand{\beq}{\begin{equation}}
\newcommand{\eeq}{\end{equation}}
\newcommand{\beqn}{\begin{eqnarray}}
\newcommand{\eeqn}{\end{eqnarray}}

\newcommand{\ga}{\mbox{${\gamma}$}}

\newcommand{\de}{\mbox{${\delta}$}}

\newcommand{\om}{\mbox{${\omega}$}}

\begin{document}

\begin{titlepage}

\vspace{1cm}

\begin{center}
{\bf \large Density of dark matter in Solar system\\ and
perihelion precession of planets}
\end{center}

\vspace{5mm}

\begin{center}
I.B. Khriplovich\footnote{khriplovich@inp.nsk.su}\\
{\em Budker Institute of Nuclear Physics\\
11 Lavrentjev pr., 630090 Novosibirsk, Russia,\\
and Novosibirsk University}
\end{center}
\bigskip

\begin{abstract}
Direct model-independent relation between the secular perihelion
precession of a planet and the density of dark matter $\rho_{\rm
\;dm}$ at its orbit is indicated, and used to deduce upper limits
on local values of $\rho_{\rm \;dm}$.

\end{abstract}

\end{titlepage}

The analysis of the perihelion precession (though not of a planet,
but of the asteroid Icarus) was for the first time used to obtain
an upper limit on the dark matter density (dmd) in the Solar
system in \cite{gr}. This limit was on the level
\begin{equation}\label{gs}
\rho_{\rm \;dm} < 10^{-16}\;{\rm g/cm^3}\,.
\end{equation}

Recently, precision EPM ephemerides were constructed in
\cite{pit1,pit2}. The analysis of thus obtained possible
corrections (i.e. of the deviations of the results of theoretical
calculations from the observational data) to the secular
perihelion precession $\de\phi$ of three planets resulted in much
stronger upper limit on the dmd in the Solar system on the level
\cite{kp}
\begin{equation}\label{kp}
\rho_{\rm \;dm} < 3 \times 10^{-19}\;{\rm g/cm^3}\,.
\end{equation}
More sophisticated analysis of the data and of possible effects
performed in [5--7] has confirmed this result, at least
qualitatively.

In the present note I wish to point out a simple relation
pertinent to the problem, as well as a quite important physical
conclusion following from this relation. The relation applies to
any spherically symmetric, but otherwise arbitrary, dmd $\rho_{\rm
\;dm}(r)$. If the orbit eccentricity $e$ for a planet is small,
the relative shift of its perihelion per period can be written as
\begin{equation}\label{r}
\frac{\de\phi}{2\pi}\, = -\,\frac{2\pi\, \rho_{\rm \;dm}(r)\,r^3}
{M_\odot}\,,
\end{equation}
Here $r$ is the radius of the (approximately circular) orbit, and
$M_\odot$ is the mass of the Sun. Formula (\ref{r}) is accurate up
to a correction on the order of $e^2$. This correction is small
for any planet of our Solar system, and can be safely ignored for
the discussed problem, at least at the present level of accuracy.

Formula (\ref{r}) can be derived, for instance, as follows. The
perturbation of the gravitational potential for a planet of mass
$m$ moving along a circular orbit of radius $r$ is
\begin{equation}\label{deu}
\de U(r) = \, k \,m\,\int^r \frac{dr_1}{r_1^2}\mu(r_1)\,,
\end{equation}
where
\[
\mu(r) = 4\pi\int_0^{r_1}\rho(r_2)r_2^2 dr_2\,
\]
is the total mass of dark matter inside a sphere of radius $r$,
and $k$ is the Newton gravitational constant (as usual, the
potential is defined up to a constant, so that the value of the
lower integration limit in (\ref{deu}) is irrelevant). Then one
computes the corrections, due to the perturbation $\de U(r)$, to
the frequency $\om_\phi$ of rotation and to the frequency $\om_r$
of radial oscillations. The difference between these corrections
multiplied by the unperturbed period of rotation gives the
perihelion shift.

Obviously, under the assumption of constant density $\rho_{\rm
\;dm}$ made in \cite{kp}, equation (\ref{r}) goes over into
formula
\begin{equation}\label{kp1}
\frac{\de\phi}{2\pi} = -\,\frac{3}{2}\,\frac{\mu(r)}{M_\odot}\,,
\end{equation}
derived therein; here and below $\mu(r)$ is the total mass of dark
matter inside the sphere of radius $r$ (we have omitted in rhs of
(\ref{kp1}) the factor $\sqrt{1-e^2}$ present in the corresponding
formula of \cite{kp}). It should be mentioned here also that a
formula equivalent to (\ref{r}), but in the special case of
constant dmd, was presented in \cite{se} (also with the factor
$\sqrt{1-e^2}$ in rhs). On the other hand, under the assumption
\beq\label{rg}
\rho_{\rm \;dm} = \rho_0 (r/r_0)^{-\gamma}
\eeq
made in \cite{fr}, equation (\ref{r}) is equivalent to relation
\begin{equation}\label{fr}
\frac{\de\phi}{2\pi} =
-\,\frac{3-\ga}{2}\,\frac{\mu(r)}{M_\odot}\,
\end{equation}
from that reference.

Let us emphasize here the following. All equations, (\ref{r}),
(\ref{kp}), and (\ref{fr}), are valid only under the assumption,
implicit or explicit, that the dark matter density $\rho_{\rm
\;dm}(r)$ is spherically-symmetric with the center coinciding with
the Sun. Such a picture looks reasonable since the typical limits
on dark matter density in the Solar system discussed in [4--7],
are much higher than typical galactic values of $\rho_{\rm \;dm}$.

Then, the assumption of constant dm density made in [4--6] in the
analysis of perihelion rotation for inner planets, corresponds to
the situation when $\rho_{\rm \;dm}$ varies at the distances much
larger than 1 astronomic unit (au). However, it is far from being
clear what could be the true scale of $\rho_{\rm \;dm}(r)$
variation. The same word of caution refers to the assumption
(\ref{rg}) with $r_0 = 1$ au and $\ga \sim 1$, so much the more
when it is applied, in line with near planets, to Jupiter, Saturn,
Uranus.

The special physical implication of formula (\ref{r}) is as
follows. The perihelion rotation is governed directly by a local
dark matter property, i. e. by its density $\rho_{\rm \;dm}(r)$ at
the planet trajectory of radius $r$, but not by its global
property, the total mass $\mu(r)$ of dark matter inside the sphere
of radius $r$. Therefore, the analysis of the observational data
for the secular perihelion precession of various planets results
in direct, model-independent upper limits on the local dmd at
various distances from the Sun corresponding to the orbit radii.
The results of this analysis are presented in Table 1. The upper
limits in the last line of the Table are derived in an obvious
simple-minded way from the numbers in the previous line.

\begin{center}
\begin{tabular}{|c|c|c|c|} \hline
 & & &  \\
&Mercury  & Earth & Mars \\
& & &  \\ \hline
& & &  \\
absolute perihelion shift, &&&\\ $''$ per century &
$-$\,0.0036\,$\pm$\,0.0050 & $-$\,0.0002\,$\pm$\,0.0004 &
0.0001\,$\pm$\,0.0005\\
& & &  \\ \hline
& & & \\
relative perihelion shift, &&&\\
$(\de\phi/2\pi)\times
10^{11}$ & $-$\,0.67\,$\pm$\,0.93  & $-$\,0.15\,$\pm$\,0.31 &  0.14\,$\pm$\,0.73\\
 & & & \\ \hline
 & & & \\
orbit radius $r$, au  & 0.39 & 1.00 & 1.52\\
& & & \\ \hline
& & & \\
&&&\\
 &1.1(1.5)$\times 10^{-17}$ &1.4(3.0)$\times
10^{-19}$ & $-$ 0.4(2.0)$\times 10^{-19}$\\
dark matter density, g/cm$^3$ & & & \\
  & $< 2.6\times 10^{-17}$ &$<
4.4\times 10^{-19}$ & $<
1.6\times 10^{-19}$\\
& & & \\ \hline
\end{tabular}

\vspace{5mm} Table 1. Dark matter density at different distances
from the Sun
\end{center}

The corresponding data from \cite{pit2} on the perihelion
precession of Venus (as well as the analogous information on far
planets, Jupiter, Saturn, Uranus) are not considered here, since
they are essentially less accurate.

Of course, the limits presented in Table 1 agree with those
derived previously. But they are obtained without extra
assumptions, in a model-independent way, and refer to various
distances from the Sun.

\begin{center}***\end{center}
I am grateful to A.A. Pomeransky for the interest to the work and
useful discussions.\\ The investigation was supported in part by
the Russian Foundation for Basic Research through Grant No.
05-02-16627.

\end{document}